\documentclass[12pt,nohyper,letterpaper]{JHEP3}         % For preprint
\usepackage{amssymb,amsfonts}
\usepackage{epsfig}
\def\be{\begin{eqnarray}}
\def\ee{\end{eqnarray}}
\newcommand{\nn}{\nonumber}
\newcommand\para{\paragraph{}}
\newcommand{\ft}[2]{{\textstyle\frac{#1}{#2}}}
\newcommand{\eqn}[1]{(\ref{#1})}

\def\Dslash{\,\,{\raise.15ex\hbox{/}\mkern-12mu D}}
\def\Dbarslash{\,\,{\raise.15ex\hbox{/}\mkern-12mu {\bar D}}}
\def\delslash{\,\,{\raise.15ex\hbox{/}\mkern-9mu \partial}}
\def\delbarslash{\,\,{\raise.15ex\hbox{/}\mkern-9mu {\bar\partial}}}
\def\pslash{\,\,{\raise.15ex\hbox{/}\mkern-9mu p}}
\def\calDslash{\,\,{\raise.15ex\hbox{/}\mkern-12mu {\cal D}}}

\newcommand{\Tr}{{\rm Tr}}

\def\lae{\mathrel{\mathop{\smash{\lower .5 ex \hbox{$\stackrel<\sim$}}}}}
\def\lae{\mathrel{\mathop{\smash{\lower .5 ex \hbox{$\stackrel>\sim$}}}}}

\title{Superconformal Vortex Strings}
\author{David Tong \\
Department of Applied Mathematics and Theoretical Physics, \\
University of Cambridge, UK\\
{\tt d.tong@damtp.cam.ac.uk}}

\abstract{We study the low-energy dynamics of semi-classical
vortex strings living above Argyres-Douglas superconformal field
theories. The worldsheet theory of the string is shown to be a
deformation of the ${\bf CP}^N$ model which flows in the infra-red
to a superconformal minimal model. The scaling dimensions of chiral
primary operators are determined and the dimensions of the associated
relevant perturbations on the worldsheet and in the four
dimensional bulk are found to agree. The vortex string thereby
provides a map between the A-series of N=2 superconformal theories
in two and four dimensions.}

\begin{document}
\pagestyle{plain} \setcounter{page}{1}
\newcounter{bean}
\baselineskip16pt

\section{Introduction}

Vortex strings provide an interesting probe of four dimensional
quantum field theories, where questions about the strongly coupled
gauge dynamics can be answered by studying the solitonic string
worldsheet \cite{sy,vstring}.

\para

In this approach one starts with a $U(N_c)$ gauge
theory, coupled to a number $N_f$ of fundamental scalar fields
$Q$, where $N_f\geq N_c$. In general, the low-energy physics of
interest is strongly coupled; let us call it Phase A. To study this 
system, the theory is first deformed by inducing a vacuum expectation value 
for $Q$. If $Q$ is made sufficiently large, and the gauge group is fully 
Higgsed, then this deformed theory will be weakly coupled. We will refer to 
this weakly coupled system as Phase B.

\para
While Phase B is amenable to semi-classical analysis, it 
appear to be an unlikely place to understand the strongly coupled 
dynamics of Phase A. However, Phase B admits vortex strings, 
stabilized by the winding of $Q$ in the plane transverse to the string.  
While the 4d bulk is weakly coupled, the low-energy 2d dynamics of the string is 
typically strongly coupled, and has been shown to capture information about 
the original Phase A of the 4d theory. The low-energy modes of interest on the 
string worldsheet arise from the embedding of the vortex in the $U(N)$ gauge group, 
and the resulting worldsheet dynamics is described by
some variant of the ${\bf CP}^{N-1}$ sigma-model \cite{vib,auzzi}. In certain 
systems, the quantum fluctuations of this 2d ${\bf CP}^{N-1}$ sigma-model 
mirror the underlying fluctuations 
of the 4d $U(N)$ non-Abelian gauge theory in Phase A. 
Indeed, analogies between 2d sigma-models and 4d gauge theories
have been studied for over 30 years: the vortex string provides a
map between the two.

\para

The programme described above was first implemented in ${\cal
N}=2$ supersymmetric theories, for which the worldsheet theory of
the vortex string has ${\cal N}=(2,2)$ supersymmetry. It was shown
in \cite{sy,vstring}, following earlier work of \cite{nick,dht},
that one may recover the Seiberg-Witten curve
\cite{sw,sw2,ho,apshap} from the worldsheet. Moreover, the exact
BPS quantum spectrum of the 2d worldsheet theory coincides with
that of the 4d gauge theory, with the quarks and W-bosons
appearing as elementary excitations of the string, while the
monopoles, which are necessarily confined in the Higgs phase,
appear as kinks on the vortex string \cite{memono}. Systems with
less supersymmetry were subsequently discussed in
\cite{other,others} where qualitative agreement between the
worldsheet and bulk theories was found.

\para

The purpose of this paper is to study a limit in which the vortex
worldsheet becomes superconformal. It is well known that there
exist special loci on the moduli space of four dimensional ${\cal
N}=2$ gauge theories where particles carrying mutually non-local
charges become massless \cite{ad,apsw,ehiy,eh}. At these
``Argyres-Douglas points", the low-energy physics is described by
a  strongly interacting superconformal field theory (SCFT). We
will examine the worldsheet theory of the vortex string associated
to this point: it is given by the ${\cal N}=(2,2)$ supersymmetric
${\bf CP}^{N-1}$ sigma-model, deformed through the addition of a
classical potential. We will see that the parameters of the
potential are tuned so that the ${\bf CP}^{N-1}$ model flows to
an interacting SCFT which we identify as the $A_{N-1}$ mimimal
model.

\para

We compare the scaling dimensions $D$ of chiral primary operators
in the 2d and 4d SCFTs. The spectrum of relevant perturbations in four
dimensions splits into two classes: those with $D<1$ and those
with $1\leq D<2$. Deformations in the former class are associated to
changing the parameters of the theory, while those in the latter class
are associated to changing vacuum expectation values (vevs) of fields
\cite{apsw}. We will show that the former
descend to chiral primary deformations on the worldsheet where their
scaling dimensions in the 2d SCFT coincide with those computed in
4d. In contrast, perturbations in the latter class take us away from
the Higgs vacuum and are not seen directly on the worldsheet.

\para

The paper is organized as follows. Section 2 deals with the bulk theory in Phase A.
We review the classical four-dimensional gauge theory of interest,
identify its superconformal point and compute the scaling
dimensions of chiral primary operators. To my knowledge, this
particular Argyres-Douglas point on the moduli space has not been
previously discussed in the literature although, as we shall see,
the resulting SCFT is not new and falls into the standard ADE
classification \cite{ehiy,eh}: we will find the $A_{2N-1}$ 4d SCFT
appearing in the moduli space of $U(N)$ gauge theory with $N$
hypermultiplets.  Section 3 deals with the worldsheet. After
reviewing how the ${\bf CP}^{N-1}$ model arises as the low-energy
dynamics of the vortex string, we identify its superconformal
point and show that the dimensions of chiral primary operators
coincide with those in the 4d bulk. We further show how motion
along the Higgs branch of the four-dimensional theory induces a
superpotential on the worldsheet and comment briefly on a novel
type of mirror symmetry of finite ${\cal N}=(2,2)$ theories.

\section{The Bulk Theory} \setcounter{section}{2}

Throughout this paper we will study ${\cal N}=2$ supersymmetric
gauge theories in four dimensions, with $U(N_c)$ gauge group and
$N_f\geq N_c$ fundamental flavors. In terms of ${\cal N}=1$
superfields, the ${\cal N}=2$ theory contains a vector multiplet
$W_\alpha$ and a chiral multiplet $\Phi$, both in the adjoint of
the gauge group. $W_\alpha$ and $\Phi$ form the ${\cal N}=2$
vector multiplet. There are further chiral multiplets $Q_i$ in the
fundamental ${\bf N_c}$ representation, and $\tilde{Q}_i$ in the
$\bar{\bf N}_c$ representation, where $i=1,\ldots, N_f$ is the
flavor index. $Q_i$ and $\tilde{Q}^i$ form the ${\cal N}=2$
hypermultiplet. We denote the complex scalar components of $\Phi$,
$Q_i$ and $\tilde{Q}_i$ by the same letter.

\para

This theory admits semi-classical vortex strings only when it
lives in the Higgs phase, in which the gauge group is
spontaneously broken by inducing a vacuum expectation value for
$Q$ (referred to as Phase B in the introduction). 
We may implement this in a manner consistent with ${\cal
N}=2$ supersymmetry by turning on a Fayet-Iliopoulos parameter
$v^2$ for the central $U(1)\subset U(N_c)$. The adjoint-valued
D-term equation then imposes $[\Phi,\Phi^\dagger]=0$, together
with
\be \sum_{i=1}^{N_f}\ Q^a_i\,(Q^\dagger)^i_b -
(\tilde{Q}^\dagger)^a_i\ \tilde{Q}^i_b\ =v^2\delta^a_b\label{d}\ee
with $a,b=1,\ldots,N_c$ the color indices.  Because the left-hand
side of \eqn{d} has rank $N_f$, while the right-hand side has rank
$N_c$, there can be solutions only when $N_f\geq N_c$ and we
restrict to this case. For $N_f<N_c$ there are no supersymmetric
vacua and, more importantly for us, no vortices. The vacuum
structure is also dictated by the superpotential, which is fixed
by ${\cal N}=2$ supersymmetry to be of the familiar form ${\cal
W}= \sum_i\,\tilde{Q}_i(\Phi-m_i)Q_i$ with $m_i$ complex mass
parameters. The resulting F-term equations are
\be \sum_{i=1}^{N_f}\,Q_i^a\,\tilde{Q}_b^i = 0\ \ \ \ , \ \ \ \
\sum_{b=1}^{N_c} \Phi^a_b\,Q_i^b = m_iQ_i^a \ \ \ \ ,\ \ \ \
\sum_{b=1}^{N_c} \tilde{Q}^i_b\,\Phi^b_a =
m_i\tilde{Q}^i_a\label{f}\ee
The supersymmetric vacuum states of the theory are given by
solutions to \eqn{d} and \eqn{f}, together with
$[\Phi,\Phi^\dagger]=0$. When $v^2=0$, there is a Coulomb branch
of vacua, parameterized by $\Phi={\rm
diag}(\phi_1,\ldots,\phi_{N})$. This Coulomb branch is the Phase A 
referred to in the introduction. In contrast, when $v^2\neq 0$, the
Coulomb branch is lifted and $\Phi$ is forced to take specific
values. If the masses $m_i$ are distinct, there are ${N_f\choose
N_c}$ isolated vacua in which $N_c$ of the $N_f$ quark fields $Q$
get an expectation value. Without loss of generality, we choose to
work with the vacuum in which the first $N_c$ flavors turn on,
\be \Phi={\rm diag}(m_1,\ldots,m_{N_c})\ \ \ ,\ \ \
Q^a_i=v\delta^a_i\ \ \ \ ,\ \ \ \ \tilde{Q}^i_a=0\label{vac}\ee
This is Phase B. The spectrum of excitations around this vacuum is gapped.
However, as the parameters are varied, new massless fields can
appear, sometimes accompanied by new, continuous, branches of
vacua. For example, when $v^2=0$, a Coulomb branch of vacua opens
up, parameterized by $\Phi$. In contrast, when some subset of the
masses coincide, a Higgs branch of vacua opens up, parameterized
by gauge invariant combinations of $Q$ and $\tilde{Q}$.

\para
Before discussing the quantum theory, let us pause briefly to
examine the pattern of symmetry breaking. As well as the $U(N_c)$
gauge symmetry, the theory also enjoys an $SU(N_f)$ flavor
symmetry. Both of these are broken spontaneously in the vacuum
\eqn{vac} in way that locks color and flavor rotations together,
\be G\cong U(N_c)\times SU(N_f)\stackrel{v}{\longrightarrow}
H\cong [U(N_c)_{\rm diag}\times U(N_f-N_c)]/U(1)\label{symm}\ee
Notice in particular that the central $U(1)\subset U(N_c)$ is
broken, providing the topology necessary to support vortex
strings. The right-hand side of \eqn{symm} is further, explictly
broken by masses $m$, which transform in the adjoint
representation of the flavor group. When $m_i\neq m_j$ for all
$i\neq j$, only the  Cartan subalgebra remains,
\be H\stackrel{m}{\longrightarrow} U(1)^{N_f-1}\
.\label{flavor}\ee

\subsection{The Superconformal Point}

In the following section, we will study the quantum dynamics of
vortex strings which exist in the classical vacuum \eqn{vac}.  
For now, we wish to study the quantum dynamics in four-dimensions. As
explained in the introduction, our interest lies ultimately not in
the Phase B vacuum \eqn{vac} --- which is weakly coupled when $v^2$ is
sufficiently large --- but instead in Phase A with $v^2=0$. This phase is defined 
by starting in \eqn{vac}, and adiabatically changing $v^2$ to zero. We 
wish to ask where on the Coulomb branch we end up. Classically, this vacuum is given by
\be \Phi={\rm diag}(m_1,\ldots,m_{N_c})\ \ \ ,\ \ \
Q^a_i=\tilde{Q}^i_a=0\label{cvac}\ee
which defines a point on the Coulomb branch. However, this vacuum
may receive quantum corrections. In general, the vacuum we want 
is the point on the Coulomb branch known as
the ``root of the baryonic Higgs phase". At this point, $N_c$ of
the $N_f$ flavors of quarks develop a massless component, ensuring
that a FI parameter $v^2$ may induce a vev for the baryon operator
$B_{1\ldots N_c}=\epsilon_{a_1\ldots a_{N_c}}Q^{a_1}_{1}\ldots
Q^{a_{N_c}}_{N_c}= v^{N_c}$, without affecting $\Phi$.
%\footnote{When the
%masses coincide, a Higgs branch of vacua emerges from this point,
%which is accordingly referred to as the ``root of the baryonic
%may only enter the Higgs phase by turning on the Fayet-Iliopoulos
%parameter $v^2$.}

\para
To determine the correct vacuum on the quantum corrected Coulomb
branch, we turn to the Seiberg-Witten curve \cite{sw,sw2}. For
$N_f<2N_c$, the curve is given by\footnote{For the $N_f=2N_c-1$
theory, it is customary to shift the masses appearing in the curve
by $m_i\rightarrow m_i+\Lambda/N_c$.}  \cite{ho,apshap}
\be y^2 =
\prod_{a=1}^{N_c}(x-\phi_a)^2-4\Lambda^{2N_c-N_f}\prod_{i=1}^{N_f}(x-m_i)
\label{curve}\ee
with $\Lambda$ the strong coupling scale of the gauge theory,
given in terms of the 4d gauge coupling $e^2$, defined at the RG
subtraction point $\mu$,
\be \Lambda^{2N_c-N_f}=\mu^{2N_c-N_f}
\exp\left(-\frac{4\pi^2}{e^2(\mu)}\right)\label{lam}\ee
The presence of $N_c$ massless quark fields provides a smoking gun
in the search for the root of the baryonic Higgs phase, for the
curve must develop a suitable degeneracy at this point. Assuming
that the first $N_c$ of the $N_f$ quarks will become massless, it
will prove notationally useful to relabel the excess masses
$\tilde{m}_i=m_{N_c+i}$ with $i=1,\ldots, N_f-N_c$. Then the root
of the baryonic Higgs phase is given by \cite{dht},
\be
\prod_{a=1}^{N_c}(x-\phi_a)=\prod_{i=1}^{N_c}(x-m_i)+\Lambda^{2N_c-N_f}
\prod_{i=1}^{N_f-N_c}(x-\tilde{m}_i)\label{root}\ee
which is to be considered as an equation for $\phi_a$ with fixed
$m_i$ and $\tilde{m}_i$. Note that in the weak coupling regime,
$m_i\gg \Lambda$, this coincides with the classical vacuum
\eqn{cvac} while, in the opposite extreme $m_i=\tilde{m}_i=0$, it
agrees with the root of the baryonic Higgs branch described in
\cite{aps}. To see that this is indeed the correct point we can
re-examine the curve, which degenerates when \eqn{root} is obeyed,
\be
y^2=\left(\prod_{i=1}^{N_c}(x-m_i)-\Lambda^{2N_c-N_f}\prod_{i=1}^{N_f-N_c}
(x-\tilde{m}_i)\right)^2\label{degcurve}\ee
signalling the presence of the desired $N$ massless quark
hypermultiplets.

\para
For fixed masses $m_i$ and $\tilde{m}_i$, the center of the vortex
string is therefore described by the point \eqn{root} on the
Coulomb branch. Our goal now is to tune the masses $m_i$, leaving
$\tilde{m}_i$ fixed, such that further states become massless. If
these carry mutually nonlocal charges with respect to the quarks,
then the resulting four dimensional theory will be superconformal.
One finds the maximally degenerate curve is given when the masses
$m_i$ satisfy
\be \prod_{i=1}^{N_c}(x-m_i)=x^{N_c}+
\Lambda^{2N_c-N_f}\prod_{i=1}^{N_f-N_c}(x-\tilde{m}_i)
\label{ad}\ee
at which point the curve is simply $y^2=x^{2N_c}$. At this point,
magnetic (or dyonic) degrees of freedom become light, joining with
the quarks to form an interacting SCFT. There are a number of ways
to see that this is indeed the case. For example, a simple
criterion was provided in \cite{apsw} which states that one gets
an interacting SCFT if extra particles become massless without
opening up new Higgs branches of vacua; one may indeed check that
no new vacuum moduli spaces appear when \eqn{ad} is satisfied.

\subsection{The Case $N_f=N_c$}

For the remainder of this section, we will focus on the simplest
case with $N_f=N_c\equiv N$ for which all the important elements
are present. We will return to the general case of $N_f>N_c$ in
section 3.3. Equation \eqn{ad} defining the superconformal point
is now
\be  \prod_{i=1}^{N}(x-m_i)=x^{N} + \Lambda^{N}\ee
which is simply solved by tuning the masses to the critical point,
\be m_k=-\exp(2\pi i k/N)\Lambda\ \ \ \ ,\ \ \ \
k=1,\ldots,N\label{masses}\ee
We would now like to identify which SCFT we have found by
computing the dimensions of chiral primary operators. This may be
achieved by expanding the Seiberg-Witten curve around the singular
point \cite{apsw}. Let us recall how this works:

\para
Our ${\cal N}=2$ theory has a classical $U(1)_R$ symmetry that
suffers an anomaly: only a $Z_{2N}$ subgroup survives
quantization. This remnant discrete symmetry is itself broken
explicitly by the masses. However, at the critical point
\eqn{masses}, superconformal invariance requires that an
accidental $U(1)_R$ symmetry is restored in the infra-red. This
enhanced symmetry is manifest in the curve which, at the singular
point,  is invariant under $U(1)_R$ with the charge assignment
$R[y]=N R[x]$. The dimensions of chiral primary operators in
${\cal N}=2$ superconformal theories satisfy $D=2I + \ft12R$,
where $R$ is the $U(1)_R$ charge and $I$ is the $SU(2)_R$ spin.
The chiral primary operators of interest deform the SCFT along the
Coulomb branch and have $I=0$. Hence $D=\ft12 R$. Expanding the
curve about the superconformal point provides a method to compute
the $R$-charge, and hence the dimensions, of all chiral primary 
perturbations. To perform this calculation, it is useful to employ
the parametrization,
\be \prod_{a=1}^{N}(x-\phi_a)=x^N+\sum_{j=1}^{N}s_j\,x^{N-j}\ \ \
,\ \ \ \prod_{i=1}^{N}(x-m_i)=x^N+\sum_{j=2}^{N}\nu_j\,x^{N-j} \ee
Notice that we have set $\nu_1=\sum_{i=1}^{N}m_i=0$, which we may
always do in a $U(N_c)$ gauge theory by a suitable shift of
$s_1=\Tr\,\Phi$. The superconformal point \eqn{masses} corresponds
to $s_j=\nu_j=0$ for $j=1,\ldots,N-1$ and $\nu_N=\Lambda^N$,
$s_N=2\Lambda^N$. Expanding about this superconformal point, we
write $\nu_j=\hat{\nu}_j$ for $j=2,\ldots, N-1$;
$s_j=\hat{\nu}_j+\hat{s}_j$ for $j=1,\ldots,N-1$;
$\nu_N=\Lambda^N+\hat{\nu}_N$ and
$s_N=2\Lambda^N+\hat{\nu}_N+\hat{s}_N$. The deformations
$\hat{\nu}_j$ shift both the masses and the expectation values and
leave us at the root of the baryonic Higgs phase \eqn{root}. In
contrast,  the deformations $\hat{s}_j$ take us away from this
locus. Expanding the curve \eqn{curve} for $N_f=N_c$ around the
singularity at $x=0$ we have
\be y^2& \approx & x^{2N}
+4\Lambda^N\sum_{j=1}^N\hat{s}_j\,x^{N-j} +
2x^N(\sum_{j=2}^N\hat{\nu}_j\,x^{N-j})+(\sum_{j=2}^N\hat{\nu}_j\,x^{N-j}
+\sum_{j=1}^N\hat{s}_j\,x^{N-j})^2 \nn\ee
To preserve the the Argyres-Douglas singularity, we must assign
relative scaling dimensions to the operators,
\be D[\hat{\nu}_j]=j\,D[x]\ \ {\rm  and}\ \
D[\hat{s}_j]=(N+j)\,D[x]\ .\ee
It remains to determine the dimension of $x$ itself, and hence the
overall normalization. This is fixed using the fact that BPS
masses are obtained by integrating the Seiberg-Witten 1-form
$\lambda_{SW}$ around closed cycles $\alpha^a$ of the curve. The
1-form is given by
\be \lambda_{SW}=\frac{1}{2\pi i} \frac{\partial P(x)}{\partial
x}\,\frac{x\,dx}{y}\ \ \ \ \ ,\ \ \ \ P(x)=\prod_{a=1}^N
(x-\phi_a)\ee
The dual scalar $\phi_D^a$, which necessarily has dimension
$D[\phi_D^i]=1$, is then obtained by  the contour integral
$\partial \phi_D^a/\partial s_b=\oint_{\alpha_a}\
\partial\lambda_{SW}/\partial s_b$. The upshot of this is that the
dimensions are constrained to obey
$D[\hat{s}_j]+(N-j+1)\,D[x]-D[y] = 1$, from which we learn the
spectrum of relevant perturbations of the SCFT,
\be D[\hat{\nu}_j]= \frac{j}{N+1} \ \ \ && \ \ \ \ j=2,\ldots,N \label{4dim}\\
D[\hat{s}_j]= \frac{N+j}{N+1} \ \ \ && \ \ \ \ j=1,\ldots,N\nn\ee
The deformations with dimensions $D<1$ are associated to varying
the mass parameters of the theory and leave us at the root of the
baryonic Higgs phase where vortices exist. As we will see shortly,
these deformations manifest themselves on the string worldsheet.
In contrast, deformations with dimension $D\geq 1$ involve only a
variation of field expectation values, and take us away from the
root of the baryonic Higgs phase where no vortex strings live. The
two types of deformations are analogous to the familiar
non-normalizable and normalizable perturbations in AdS/CFT. As
explained in \cite{apsw}, it is a general feature of 4d ${\cal
N}=2$ SCFTs that these two types of relevant parameters come in
pairs satisfying
\be D[\hat{\nu}_j]+D[\hat{s}_{N-j+2}]=2\ ,\label{doubling}\ee
The mass parameters $m_i$ are associated to the $U(1)^{N-1}$
flavor symmetry \eqn{flavor}: once these symmetries are weakly gauged, the
masses appear as background expectation values.
The fact that $D[\hat{\nu}_j^{1/j}]\neq 1$ implies that this
flavor symmetry does not act in the SCFT, but rather couples,
after weak gauging, through irrelevant interactions. Giving an
expectation value to turn on the masses then deforms the
SCFT by a relevant operator which, from the pairing \eqn{doubling},
takes the form
\be \delta {\cal L} = \sum_{j=2}^N \nu_j\,\int d^4\theta\ S_{N-j+2}\ .
\label{n2def}\ee
where $S_j$ is the ${\cal N}=2$ superfield containing $s_j$ as its lowest
component and $\int d^4\theta$ denotes integration over one half of ${\cal N}=2$
superspace. In contrast, the dimension of $D[s_1]=1$ implies that
the associated singlet mass $\Tr\,\Phi$ couples to a current
--- identified with $U(1)_B$ in the $SU(N_c)$ theory
--- which gives rise to conserved charges within the SCFT.

\para

We note in passing that our SCFT lies at a  different point than
those usually discussed in the literature. For example, in
\cite{ehiy,eh} one sets all the masses equal, $m_i=m$, and
subsequently adjusts $m$ to find further massless particles. Here,
however, we have set $\sum m_i=0$ and sought the superconformal
point lying at the root of the Higgs phase \eqn{root}.
Nonetheless, the spectrum of chiral primary operators that we have
found falls within the categorization presented in \cite{ehiy,eh};
indeed, the SCFT at the root of the $U(N)$ baryonic Higgs phase is
the same as the one within the pure $SU(2N)$ super Yang-Mills
theory. This is the $A_{2N-1}$ series of 4d ${\cal N}=2$ SCFTs.

\section{The Worldsheet Theory}
\setcounter{section}{3}

In this section we return to Phase B, described by the
classical vacuum \eqn{vac}, in which the gauge group is fully 
Higgsed by the expectation value of $Q$. We will construct an infinite,
straight vortex string in this background and study its low-energy
dynamics. We will show that precisely when the masses are tuned to
the Argyres-Douglas point, the worldsheet theory will also flow to
a SCFT which we identify as the ${\cal N}=(2,2)$ minimal model.

\para
In the Higgs vacuum, the topology $\Pi_1(G/H)\cong {\bf Z}$ of the
symmetry breaking described in \eqn{symm} supports vortex strings,
stabilized by the phase of $Q^i_a$ winding in the plane transverse
to the string. Straight, infinite vortex strings stretched in,
say,  the $x^3$ direction are BPS objects described by solutions
to the classical non-Abelian Bogomolnyi equations,
\be {\cal D}_1 Q_i=i{\cal D}_2Q_i\ \ \ , \ \ \ F_{12}=e^2 (\sum_i
Q_i Q_i^\dagger -v^2)\label{bog}\ee
Here $e^2$ is the gauge coupling constant. Both $\Phi$ and
$\tilde{Q}$ remain in their classical vacuum state \eqn{vac} in
the vortex solution. Equations \eqn{bog} are the non-Abelian
generalization of the vortex equations appearing in the Abelian
Higgs model. The solutions describe strings of tension $T=2\pi
v^2$ and width\footnote{Throughout this paper we have not distinguished 
the four-dimensional $U(1)$ gauge coupling from the $SU(N_c)$ gauge coupling. 
The width $L$ of the vortex string is determined by the Abelian coupling 
constant.} $L=1/ev$. As we now review, 
the non-Abelian embedding endows the vortex with interesting dynamics.

\subsection{The Classical Worldsheet: $N_f=N_c$}

The low-energy dynamics of vortex strings always includes two
Goldstone modes associated to their transverse fluctuations. More
important for us will be further massless (or light) modes on the
worldsheet that arise from the embedding of the vortex in the
non-Abelian gauge group. In this section we review the internal
modes of the vortex, restricting to the $N_f=N_c\equiv N$ theory.
We will return to the general case of $N_f>N_c$ at the end of the
paper. A more complete review of these solitons can be found in
\cite{tasi}.

\para
We start by describing the vortex worldsheet dynamics when the
masses $m_i=0$. Suppose we have a solution $(q,a)$  to
the Abelian $U(1)$ vortex equations. Then we may always construct
a solution to the non-Abelian vortex equations by an embedding in
the upper-left-hand corner,
\be   Q^a_i=\left(\begin{array}{ccc} q \  & & \\ & \ddots &
\\ &  & \ v\end{array}\right) \ \ \ \ ,\ \ \ \ \
A^a_b=\left(\begin{array}{ccc} a \ & & \\
& \ddots & \\ & &  \ 0\end{array}\right) \label{solcambell}\ee
The vacuum state of the 4d theory has a surviving $SU(N)_{\rm
diag}$ symmetry, which is the diagonal combination of a gauge and
flavor rotation. This acts on the solution \eqn{solcambell} as
$Q\rightarrow UQU^\dagger$ and $A\rightarrow
UAU^\dagger-i(\partial U)U^\dagger$ to provide further Goldstone
modes on the worldsheet. Dividing by the stabilizing group, the
internal low-energy dynamics of the vortex is described by a
$d=1+1$ sigma-model with target space \cite{vib,auzzi}
\be SU(N)_{\rm diag}/SU(N-1)\times U(1) \cong {\bf CP}^{N-1} \ee
The vortex is 1/2-BPS in the ${\cal N}=2$ 4d gauge theory,
ensuring that 2d worldsheet theory has ${\cal N}=(2,2)$
supersymmetry, with the fermi zero modes of the string providing
the worldsheet superpartners.

\para
There is a way to write the supersymmetric ${\bf CP}^{N-1}$
sigma-model in terms of an ${\cal N}=(2,2)$ supersymmetric $U(1)$
gauge theory that will prove useful in solving for the quantum
dynamics \cite{dadda,beaut,phases}. Consider an auxiliary  $U(1)$
field strength on the worldsheet, living in a twisted chiral
multiplet $\Sigma$, whose lowest component we denote as $\sigma$.
The gauge field couples to $N$ chiral multiplets $\Psi_i$, each
with charge $+1$. The lowest components of $\Psi_i$ will be
denoted as $\psi_i$ and play the role of homogeneous coordinates
on ${\bf CP}^{N-1}$. The potential energy of the worldsheet theory
is a sum of F-terms and the D-term
\be V_{2d}= \sum_{i=1}^N|\sigma|^2|\psi_i|^2+\frac{g^2}{2}
(\sum_{i=1}^N|\psi_i|^2-r)^2\ee
Here $g^2$ is a gauge coupling on the worldsheet which is
irrelevant for the infra-red quantum dynamics at energies $E\ll
g$. At low-energies, the D-term restricts to $\sum_i|\psi_i|^2=r$.
After dividing by $U(1)$ gauge transformations $\psi_i\rightarrow
e^{i\alpha}\psi_i$, the gauge theory reduces to the sigma-model
with target space ${\bf CP}^{N-1}$.  The worldsheet FI parameter
$r$ determines the size of the ${\bf CP}^{N-1}$ target space and,
for the vortex moduli space, is given in terms of the 4d gauge
coupling \cite{vib},
\be r=\frac{2\pi}{e^2}\ee
The 4d theta angle also descends to a 2d theta angle on the
worldsheet \cite{vstring,others}.

\para
So far we have discussed the theory with vanishing masses $m_i=0$.
How do non-zero masses change the worldsheet dynamics of the
vortex string? The answer was given in \cite{memono,sy,vstring}.
The masses break the surviving symmetry group $SU(N)_{\rm
diag}\rightarrow U(1)^{N-1}$ and the associated worldsheet
Goldstone modes are lifted. Of the ${\bf CP}^{N-1}$ moduli space
of solutions, only $N$ isolated solutions remain. These correspond
to the Abelian vortex $(q, a)$ embedded in the $N$
different diagonal elements of $Q$ and $A$. From the perspective
of the worldsheet theory, the complex masses $m_i$ in 4d can be
shown to induce twisted masses $m_i$ \cite{hh} in 2d, so that the worldsheet
potential energy reads
\be
V_{2d}=\sum_{i=1}^N|\sigma-m_i|^2|\psi_i|^2+\frac{g^2}{2}(\sum_{i=1}
|\psi_i|^2-r)^2\ee
As anticipated, the ${\bf CP}^{N-1}$ target space is lifted,
leaving behind $N$ isolated vacua of the vortex worldsheet given
by $\sigma=m_i$ and $|\psi_j|^2=r\,\delta_{ij}$, for
$i=1\ldots,N$. Kinks in the vortex string, interpolating between
these different worldsheet vacua, are confined 4d monopoles
\cite{memono}.

\subsection{The Superconformal Worldsheet: $N_f=N_c$}

In summary, the classical low-energy dynamics of the vortex string
in the theory with $N_f=N_c$ flavors is described  by the ${\cal
N}=(2,2)$ ${\bf CP}^{N_c-1}$ sigma model, with the 4d masses $m_i$
inducing a classical potential over the target space. In this
section we study the quantum dynamics of this theory. Typically,
the ${\bf CP}^{N-1}$ sigma-model has a mass gap. However, we will
show that once the masses are tuned so that the 4d theory lies at
the Argyres-Douglas point, the ${\bf CP}^{N-1}$ sigma model flows
to an interacting SCFT which we  identify as the $A_{N-1}$ minimal
model.

\para

The quantum effective action for the mass deformed ${\bf
CP}^{N-1}$ model was studied in \cite{hh,nick,svz}.
Following \cite{phases}, one integrates out the charged 2d chiral
multiplets $\Psi_a$ to find an effective twisted superpotential
$\tilde{\cal W}$ for the field strength $\Sigma$,
\be \tilde{\cal W} = -\frac{1}{2\pi}\sum_{i=1}^{N}
(\Sigma-m_i)\left[\log\left(\frac{\Sigma-m_i}{\mu}\right)-1\right]
- t\Sigma\label{wtilde}\ee
where $\mu$ is the RG subtraction point. The 2d complexified FI
parameter $t=r+i\theta$ runs under RG flow and is exchanged for
the invariant dynamical scale $\Lambda=\mu \exp(-2\pi t/N)$.
Because the 2d FI parameter is related to the 4d gauge coupling
through $r=2\pi/e^2$, the strong coupling scale $\Lambda$ on the
worldsheet coincides with the 4d strong coupling scale \eqn{lam}.

\para
The worldsheet theory has $N$ vacuum states, given by the critical
points of $\tilde{\cal W}$,
\be \prod_{i=1}^{N_c}(\sigma-m_i)-\Lambda^N = 0\label{dooby}\ee
For large mass differences, so $|m_i-m_j|\gg |\Lambda|$, these
vacua coincide with the classical vacua $\sigma = m_i$. In the
opposite regime, $m_i=0$, the $N$ vacua descend to the strong
coupling scale $\sigma = \omega \Lambda$, where $\omega^N=1$.
Comparing the Seiberg-Witten curve at the root of the baryonic
Higgs phase \eqn{degcurve} to the twisted superpotential
\eqn{wtilde}, we see that, for $N_f=N_c$, the former may be
written as $y^2= (\partial\tilde{W}(x)/\partial x )^2$. This is
the statement that the worldsheet of the vortex string captures
the Seiberg-Witten curve.

\para

Let us now tune the masses so that the 4d theory sits at the
Argyres-Douglas point identified in the previous section.
Following \eqn{masses}, we set $m_k=- \exp(2\pi i k/N)\Lambda$ and
examine the consequences for the worldsheet. The importance of
this point was stressed in \cite{nick,svz}. The $N$ critical points
\eqn{dooby} merge at $\sigma=0$, ensuring that the kinks
interpolating between different vacuum states become massless.
This reflects the behavior of monopoles in the underlying
Arygres-Douglas point because, in the Higgs phase, monopoles are
confined, trapped to live on the vortex string where they appear
as kinks. It was conjectured in \cite{svz} that the 2d theory
becomes an interacting SCFT at the point \eqn{masses}. To see that
this is indeed the case, we expand the twisted superpotential
\eqn{wtilde} at this point  for small $\sigma/\Lambda$ to find
\be \tilde{\cal W} = c_0 \ \frac{\Sigma^{N+1}}{\Lambda^N}+ \ldots
\label{an} \ee
where $c_0$ is an overall normalization and $\ldots$ refer to
irrelevant operators. We see that the familiar logarithms of the
${\bf CP}^{N-1}$ sigma-model are replaced by a polynomial
Landau-Ginzburg model.  The K\"aher potential of the theory is
unknown at this strongly coupled point. However, this is
unimportant because, while the superpotential is protected by
non-renormalization theorems, the K\"ahler potential is expected
to adjust itself under RG flow so that the theory flows to an
interacting ${\cal N}=(2,2)$ SCFT which is identified with the
$A_{N-1}$ minimal model \cite{martinec,vw,minimal}. The central
charge of this 2d SCFT is
\be {c}=3-\frac{6}{N+1}\ .\ee
Representation theory of the ${\cal N}=(2,2)$ superconformal
algebra relates the dimension $D$ of chiral primary operators to
the charge $R$ under the $U(1)_R$ symmetry: $D=\ft12 R$. The ${\bf
CP}^{N-1}$ sigma-model has a classical $U(1)_R$ symmetry which is
anomalous in the quantum theory, with only a ${\bf Z}_{2N}$
subgroup surviving. This surviving discrete group is further
broken explicitly by generic twisted masses $m_i$. However, at the
critical point in parameter space, where the theory is governed by
\eqn{an}, an accidental $U(1)_R$ symmetry is restored in the
infra-red, as required by superconformal invariance. This mirrors
the story for the $U(1)_R$ symmetry in four dimensions; indeed,
the worldsheet $U(1)_R$ is inherited from the 4d $U(1)_R$.

\para
Since the twisted superpotential necessarily has R-charge 2, the
R-charge of the twisted chiral multiplet $\Sigma$ is given by
$R[\Sigma]=2/(N+1)$. The spectrum of chiral primary operators
therefore have dimensions $D_j=j/(N+1)$ where $j=1,\ldots,N-1$.
(The addition of the operator $\Sigma^{N}$ with $j=N$ is redundant
since it may be absorbed by a constant shift of $\Sigma$). We may
identify each of these relevant deformations in terms of the mass
parameters $m_i$. To do this, rewrite
\be \prod_{i=1}^N(\sigma-m_i)=\sigma^N +\sum_{j=2}^{N-1}\nu_j\,
\sigma^{N-j}\ee
where there is no $\nu_1$ since we have chosen
$\sum_{i=1}^Nm_i=0$. The conformal point \eqn{masses} in parameter
space corresponds to $\nu_j=0$ for $j=1,\ldots, N-1$ and
$\nu_N=\Lambda^N$. We expand the twisted superpotential
\eqn{wtilde} about this point, writing $\nu_j=\hat{\nu}_j$ for
$j=1,\ldots,N-1$ and $\nu_N=\Lambda_N + \hat{\nu}_N$, to find
\be \delta{\tilde {\cal W}}=\sum_{j=2}^N\,c_j\,\hat{\nu}_j\,\Sigma^{N-j+1}+\ldots\ee
which is to be compared to \eqn{n2def}, giving the map between bulk and worldsheet
chiral operators: $\Sigma^{N-j+1} \leftrightarrow S_{N-j+2}$. The dimensions of
the relevant perturbations are
\be D[\hat{\nu}_j]=\frac{j}{N+1}\ \ \ \ \
j=2,\ldots,N\label{2dim}\ee
in agreement with the four-dimensional result \eqn{4dim}.

\para

The vortex string thus provides a map between the $A_{2N-1}$
series of 4d ${\cal N}=2$ SCFTs, and the $A_{N-1}$ series of 2d
${\cal N}=(2,2)$ SCFTs. Although only one half of the 4d relevant
operators are realized on the worldsheet (those that leave us at
the root of the baryonic Higgs branch), the general feature
\eqn{doubling} of 4d SCFTs ensures that we can reconstruct the
full spectrum of relevant operators from the worldsheet.

\para

Relationships between 4d SCFTs and 2d minimal models have been
described previously. In particular, the spectrum of BPS states in
the vicinity of an Argyres-Douglas point was shown to bear many
similarities to massive deformations of 2d SCFTs \cite{vaf}. The
vortex string provides a rationale for this correspondence, with
the 4d BPS states mapping to the 2d BPS states.

\subsection{Generalization to $N_f>N_c$}

So far we have examined the superconformal point only for 
$N_f=N_c$. We now briefly discuss the generalization to $N_c< N_f
\leq 2N_c-1$ flavors. For distinct masses, there are ${N_f\choose
N_c}$ different roots of the baryonic Higgs phase. We choose to
work with the root which classically corresponds to the vacuum
$\Phi={\rm diag}(m_1,\ldots,m_{N_c})$ and, as in section 2, we
relabel the $(N_f-N_c)$ remaining masses as
$\tilde{m}_i=m_{N_c+i}$.

\para

It is a simple exercise to expand the curve \eqn{degcurve} about
the superconformal point \eqn{ad} to extract the dimensions of
chiral primary operators in the four dimensional SCFT with
$N_f>N_c$. For generic non-zero masses $\tilde{m}_i$ one finds
that the singularity is unaltered, corresponding once again to a
SCFT with scaling dimensions \eqn{4dim}. The excess masses
$\tilde{m}_i$ in this case are irrelevant deformations. However,
this changes when some of the masses $\tilde{m}_i$ vanish. In this
situation the singularity is partially resolved. Consider the
extreme case $\tilde{m}_i=0$ for all $i=1,\ldots, N_f-N_c$.
Expanding the curve about the superconformal point \eqn{ad} now
gives,
\be y^2& \approx & x^{2N_c}
+4\Lambda^{2N_c-N_f}x^{N_f-N_c}\sum_{j=1}^{N_c}\hat{s}_j\,x^{N_c-j}
+ 2x^{N_c} (\sum_{j=2}^{N_c}\hat{\nu}_j\,x^{N_c-j})
+(\sum_{j=2}^{N_c}\hat{\nu}_j\,x^{N_c-j})^2+\ldots \nn\ee
where $\ldots$ are irrelevant terms. The relative scaling
dimensions of the various perturbations are now given by
$D[\nu_j]=j[x]$ and $D[s_j]=(2N_c-N_f+j)[x]$. The overall
normalization remains as before, giving us the dimensions
\be D[\hat{\nu}_j]=\frac{j}{2N_c-N_f+1}\ \ \ ,\ \ \ D[\hat{s}_j] =
\frac{2N_c-N_f+j}{2N_c-N_f+1}\label{4dnew}\ee
We will now show how this behavior is captured by the worldsheet.
Vortex strings in the theory with $N_f> N_c$ have a rather
different property from those in the $N_f=N_c$ theory: their scale
size is a collective coordinate. (See \cite{semi2} for a review).
An effective dynamics for the string was proposed in \cite{vib}.
It is an ${\cal N}=(2,2)$ supersymmetric $U(1)$ gauge theory with $N_c$ chiral
multiplets $\Psi_i$ of charge $+1$ and twisted mass $m_i$. There
are a further $(N_f-N_c)$ chiral multiplets $\tilde{\Psi}_j$ of
charge $-1$ and twisted mass $\tilde{m}_j$. The scalar potential
on the worldsheet is given by
\be
V_{2d}=\sum_{i=1}^{N_c}|\,\sigma-m_i|^2|\psi_i|^2+\sum_{i=1}^{N_f-N_c}
|\,\sigma-\tilde{m}_i|^2|\tilde{\psi}_i|^2+\frac{g^2}{2}(\sum_{i=1}^{N_c}
|\psi_i|^2 - \sum_{j=1}^{N_f-N_f}|\tilde{\psi}_j|^2-r)^2\nn\ee
When the masses vanish, $m_i=\tilde{m}_i=0$, the extra modes
$\tilde{\psi}_i$ provide the worldsheet theory with a non-compact
moduli space of vacua. This non-compact direction corresponds to
the scaling mode of the vortex. For generic values of the masses,
this worldsheet theory was shown to share its BPS spectrum with
the 4d theory in which the vortex lives \cite{dht,vstring}. Here we
examine this theory at the superconformal point.

\para
Integrating out the chiral multiplets, the effective worldsheet
theory is governed by the twisted superpotential,
\be \tilde{\cal
W}=-\frac{1}{2\pi}\sum_{i=1}^{N_c}(\Sigma-m_i)\left[\log\left(
\frac{\Sigma-m_i}{\mu}\right)-1\right]
+\frac{1}{2\pi}\sum_{j=1}^{N_f-N_c}
(\Sigma-\tilde{m}_j)\left[\log\left(\frac{\Sigma-\tilde{m}_j}{\mu}
\right)-1\right]-t\Sigma\nn\ee
The superconformal point on the worldsheet occurs when all
critical points coincide,
\be
\prod_{i=1}^{N_c}(\sigma-m_i)=\sigma^{N_c}+\Lambda^{2N_c-N_f}\prod_{j=1}^{N_f-N_c}
(\sigma-\tilde{m}_j)\ee
which is to be understood as an equation for the masses $m_i$ for
fixed $\tilde{m}_j$. Notice that, as expected, the equation
coincides with the four dimensional superconformal point defined
in \eqn{ad}. The nature of the worldsheet SCFT depends on the
masses $\tilde{m}_i$. For $\tilde{m}_i\neq 0$, expanding the
twisted superpotential about the superconformal point gives
\be \tilde{\cal W}\sim
\frac{\Sigma^{N+1}}{\Lambda^{2N_c-N_f}\prod_j\tilde{m}_j} \ ,\ee
which we recognize once again as the $A_{N-1}$ ${\cal N}=(2,2)$
SCFT. However, when $\tilde{m}_j=0$ for some $j$, the nature of
the superconformal point changes. On a technical level this occurs
because we may no longer expand $\tilde{W}$ in
$\sigma/\tilde{m}_j$. Consider the extreme case when
$\tilde{m}_j=0$ for all $j=1,\ldots, N_f-N_c$. It is simple to
repeat the computation above to find worldsheet superpotential
\be \tilde{\cal W}\sim
\frac{\Sigma^{2N_c-N_f+1}}{\Lambda^{2N_c-N_f}}\ee
corresponding to a reduced $A_{2N_c-N_f-1}$ SCFT. The dimensions
of relevant perturbations are now given by
\be
D[\hat{\nu}_j]=\frac{j}{2N_c-N_f+1}\label{2dnew}\ee
in agreement with the four dimensional theory \eqn{4dnew}. Before
moving on, we pause to note that the validity our starting
worldsheet theory is in some doubt in the case $\tilde{m}_j=0$. A
classical infra-red divergence means that the scaling modes of the
vortex string are non-normalizable \cite{ward,ls}, a fact that is
not obviously captured in the worldsheet theory described above.
For this reason, one might expect the worldsheet theory to be
valid only when $\tilde{m}_i\neq0$, which ensures that the the
infra-red divergence is rendered finite \cite{semi}. The result
\eqn{2dnew} shows that, at the superconformal point, the proposed
worldsheet theory dynamically freezes the scaling modes when
$\tilde{m}_i=0$. This, coupled with the  resulting agreement with
the four-dimensional SCFT, suggests that the worldsheet theory
continues to capture the quantum physics of the vortex string even
when $\tilde{m}_i=0$.

\subsection{Moving on the Higgs branch}

When $m_i=\tilde{m}_i=0$, the four dimensional theory with
$N_f>N_c$ has a Higgs branch of vacua of complex dimension
$N_c(N_f-N_c)$. A gauge invariant description of this branch
is provided by expectation values for the meson and baryon operators
\be M^j_i=\tilde{Q}^j_aQ^a_i\ \ \ ,\ \ \ B_{i_1\ldots
i_{N_c}}=\epsilon_{a_1\ldots a_{N_c}}Q_{i_1}^{a_1}\ldots
Q_{i_{N_c}}^{a_{N_c}}\ \ \ ,\ \ \ \tilde{B}^{i_1\ldots
i_{N_c}}=\epsilon^{a_1\ldots a_{N_c}}\tilde{Q}^{i_1}_{a_1}\ldots
\tilde{Q}^{i_{N_c}}_{a_{N_c}}\nn\ee
These are not all independent, but satisfy a number of polynomial
relations which must be imposed, together with the D-term \eqn{d}
and F-term equations \eqn{f}, to describe the Higgs branch in the
gauge invariant fashion --- see \cite{aps} for more details. The
presence of the FI parameter in the D-term \eqn{d} deforms, but
does not lift, the Higgs branch\footnote{In the language of
\cite{aps}, the non-baryonic branch {\it is} lifted by the FI
parameter, while the baryonic branch survives, deformed.}.

\para
Let us return momentarily to a description of the Higgs branch in
terms of the gauge non-invariant fields. A combination of gauge
and flavor rotations allows the general solution to the D-term
\eqn{d} and F-term \eqn{f} equations to be put in the form
\cite{aps}
\be Q^a_i= q^a\,\delta^a_i \ \ \ ,\ \ \ \ \tilde{Q}^i_a=
\tilde{q}_a\, \delta^{i}_{a+N_c} \ \ \ \ \ \mbox{no sum on
$a$}\label{point}\ee
subject to  $|q^a|^2-|\tilde{q}_a|^2=v^2$ for each
$a=1,\ldots,N_c$. Note that $\tilde{q}_a=0$ for $a> N_f-N_c$, so
this parametrization describes a $(N_f-N_c)$ dimensional slice of
the Higgs branch. To identify a point on the Higgs branch of
the form \eqn{point}, it is sufficient to give only the values of
the meson field $M^i_j$, whose non-vanishing components may be written
in the form of an $N_c\times (N_f-N_c)$ matrix,
\be \hat{M}^j_i = M^{j+N_c}_i\ \ \ \ \ i=1,\ldots,N_c\ ,\
j=1,\ldots N_f-N_c\ee
After this small digression, we now return to the vortex
worldsheet. So far we have discussed the theory on the vortex only
at a special point \eqn{vac}, which we may call the origin of the
Higgs branch. It is defined in terms of the gauge invariant fields
by
\be {\rm Origin:}\ \ M=\tilde{B}=0\ \ {\rm  and}\  \ B_{1\ldots N_c}=v^{N_c}
\label{rogin}\ee
with all other
components of $B$ vanishing. Here we would like to ask how the
vortex worldsheet theory responds to motion in the Higgs branch.
We will show that sitting on a point in the Higgs branch specified
by $\hat{M}$ induces a gauge invariant superpotential on the
vortex string worldsheet,
\be {\cal W}\sim \hat{M}^j_i\,\tilde{\Psi}_j\Psi^i\label{sup}\ee
This superpotential partially lifts the vortex moduli space. When
$\hat{M}$ is of maximal rank $(N_f-N_c)$, the surviving vortex
moduli space is ${\bf CP}^{2N_c-N_f-1}$. This reduction from $N_c$
to $(2N_c-N_f)$ is compatible with the $A_{2N_c-N_f-1}$ SCFT we
found in the previous section when $\tilde{m}_j=0$. A relationship
between the 4d Higgs branch and 2d complex masses is also suggested
by the brane picture \cite{hh}.

\para
To see that \eqn{sup} is correct, we return to the Bogomolnyi
equations for the vortex. The equations \eqn{bog} were derived
under the assumption that $\tilde{Q}=0$. Relaxing this condition,
the full Bogomolnyi equations are given by
\be {\cal D}_1 Q_i=i{\cal D}_2Q_i\ \ , \ \ {\cal D}_1
\tilde{Q}^i=i{\cal D}_2\tilde{Q}^i \ \  ,\ \  \frac{1}{e^2}F_{12}=\sum_i
(Q_i Q_i^\dagger - \tilde{Q}^{i\dagger}
\tilde{Q}^i)-v^2\label{bog2}\ee
together with the F-term condition
\be \sum_{i=1}^{N_f}Q^a_i\tilde{Q}_b^i=0\ . \label{bogf}\ee
When $\tilde{Q}$ has a vev, the space of solutions to these
equations is reduced compared to the case where we can set
$\tilde{Q}=0$. The troublesome equation is the second in
\eqn{bog2}. Components of $\tilde{Q}$ which have an expectation
value have no non-trivial solutions to this equation. In the
Abelian case this follows from the fact that there is no
holomorphic line bundle of negative degree. (It may also be seen
through a direct study of the Bogomolnyi equations \cite{penin,davis}). Let
us see the effect in the non-Abelian case. The collective coordinates
$\psi^a$, which provide homogeneous coordinates on ${\bf CP}^{N-1}$
tell us how the Abelian gauge
potential $a(x^1,x^2)$ describing a vortex profile is
embedded in the non-Abelian gauge group. The dictionary is
\be A^a_b(x^1,x^2) = \psi^a\bar{\psi}_b\ a(x^1,x^2)\ee
Writing $z=x^1+ix^2$, the equation for $\tilde{Q}$ reads
\be ({\cal D}_z\tilde{Q}^i)_a=\partial_z \tilde{Q}^i_a  +
ia_z\,\bar{\psi}_a\sum_{b=1}^{N_c}{\psi^b}\,\tilde{Q}^i_b=0\ee
This can be satisfied trivially by $\partial \tilde{Q}=0$ only if
the vortex sits inside the $U(N_c)$ gauge group in such a way that
$\sum_b\psi^b\,\tilde{Q}^i_b=0$.  At the point \eqn{point} on the
Higgs branch, this means that $\psi^a=0$ for all $a=1,\ldots,N_c$ such
that $\tilde{q}_a\neq 0$. In terms of the gauge invariant meson
observables, this condition can be re-expressed as
\be \sum_{i=1}^{N_c}\,\hat{M}^j_i \psi^i =  0 \ \ \ \mbox{for each
$j=1,\ldots, N_f-N_c$}\ee
which is indeed a subset of the restrictions that arises from the worldsheet
superpotential \eqn{sup}:  $\partial W/\partial \tilde{\psi}_j=0$.
The remaining restrictions arising from the worldsheet superpotential are
given by,
\be \frac{\partial{\cal W}}{\partial
\psi^i}=\sum_j\hat{M}^j_i\tilde{\psi}_j=0 \ \ \ \mbox{for each
$i=1,\ldots, N_c$}\label{last}\ee
These conditions remove the scaling modes $\tilde{\psi}_j$ of the vortex
string. Let us now see that these too are implied by the four-dimensional
equations of motion. As we mentioned in the last section, these scaling modes
can be traced to the presence of the excess scalar fields
$Q_{i+N_c}$, $i=1,\ldots,N_f-N_c$, that do not gain an expectation
value. At the origin of the Higgs branch \eqn{rogin} these
fields have a profile of the form
\be Q^a_{i+N_c}(x^1,x^2)=\psi^a\tilde{\psi}_{i}\,
\tilde{q}(x^1,x^2;|\tilde{\psi}_i|)\ee
where $\tilde{q}$ is the profile of an Abelian scalar
field, of the type discussed in \cite{semi2}, and
satisfies the boundary conditions
$\tilde{q}\rightarrow 0$ as $x\rightarrow \infty$.
This condition holds at the origin of the Higgs branch \eqn{rogin}.
However, once we move into the interior of Higgs branch, and the
mesonic field $\hat{M}$ is non-vanishing, these modes fall foul of
the F-term condition \eqn{bogf}. Equation  \eqn{last} imposes the requirement
of the F-term on the worldsheet.

\para
In summary, we have shown that the space of solutions to the
vortex equations \eqn{bog2} and \eqn{bogf} at a non-trivial point
on the Higgs branch is given by the zero set of the superpotential
\eqn{sup}. The remaining solutions are $1/2$-BPS, requiring that
the worldsheet theory has a vacuum preserving ${\cal N}=(2,2)$
supersymmetry. Happily, for $N_f<2N_c$, it does.

\subsection{A Comment on S-Duality and Mirror Symmetry}

So far we have focussed only on asymptotically free theories with
$N_f<2N_c$. Here we make some (very) brief remarks about the scale
invariant theory with $N_f=2N_c$. The complex gauge coupling
$\tau=2\pi i/e^2-\theta$ is an exactly marginal parameter of the
theory. The Seiberg-Witten curve is given by \cite{apshap}
\be
y^2=\prod_{a=1}^{N_c}(x-\phi_a)^2+h(q)(h(q)+2)\prod_{i=1}^{N_f}
(x-h(q)m_S-m_i)\label{sicurve}\ee
where $q=e^{2\pi i \tau}$ at weak coupling but may, in general,
receive instanton corrections. The definition of the modular
function $h(q)$ can be found in \cite{sw2,apshap}, while the
singlet mass is defined by  $m_S=\sum_im_i/N_f$. The modular
properties of the curve imply a $\Gamma_0(2)$ duality of the field
theory, where $\Gamma_0(2)$ is the subset of $SL(2,{\bf Z})$
matrices with even upper off-diagonal entry.

\para
The corresponding vortex theory is a $U(1)$ gauge theory with
$N_c$ chiral multiplets of charge $+1$ and a further $N_c$ chiral
multiplets of charge $-1$. It is  scale invariant, with the
complex FI parameter $t=r+i\theta\equiv -i \tau$ an exactly marginal
parameter.

%The exact twisted superpotential is given by
%%
%\be \tilde{\cal
%W}=-\frac{1}{2\pi}\sum_{i=1}^{N_c}(\Sigma-m_i)\,\log\left(\frac{\Sigma-m_i}{\mu}
%\right) +\frac{1}{2\pi}\sum_{i=N_c+1}^{2N_c}(\Sigma-m_i)\,\log
%\left(\frac{\Sigma-m_i}{\mu}\right) -t\Sigma \nn\ee
%%
%At first glance, the worldsheet no longer appears to capture the
%Seiberg-Witten curve at the root of the baryonic Higgs branch.
%However, the exact low-energy solutions parameterized by curves
%are rife with ambiguities due to possible field definitions
%associated to instanton effects --- see, for example
%\cite{dkm,pelland}. It was shown in \cite{pelland} that the curve
%\eqn{sicurve} can indeed be written in a way which coincides with
%$y^2=(\partial \tilde{\cal W}(x)/\partial x)^2$ at the root of the
%baryonic Higgs branch. (A related form of the curve also appears
%naturally in the M-theory construction \cite{mtheory}).

\para

The modular properties of the four-dimensional theory strongly
suggest that there is a similar duality group at play in the two
dimensional theory, interchanging kinks and elementary
excitations. Dualities of this form are, of course, familiar in
two dimensions \cite{coleman} although, to my knowledge, the exact
structure of the duality in the present system has not been worked
out. Although the duality involves an inversion of the K\"ahler
class of the target space, reminiscent of T-duality, it appears to
differ from the mirror symmetry of \cite{hv}. It would be
interesting to explore this system further.

\subsection*{Acknowledgement} My thanks to Allan Adams, Sean
Hartnoll, Anton Kapustin, Eliezer Rabinovici, and especially to
Nick Dorey and Kentaro Hori, for helping me join the dots. 
I'm supported by the Royal Society

\end{document}